\documentclass[twocolumn,10pt]{article}
\usepackage{amsmath,amssymb}

\font\block=msbm10
\def\C{\hbox{\block\char'0103}}
\def\Z{\hbox{\block\char'0132}}

\font\gotic=eufm10
\def\g{\hbox{\gotic\char'0147}}
\def\l{\hbox{\gotic\char'0154}}
\def\s{\hbox{\gotic\char'0163}}

\font\new=eusm10
\def\A{\hbox{\new\char'0101}}
\def\P{\hbox{\new\char'0120}}

\makeatletter \oddsidemargin-.25in \evensidemargin-.25in
\makeatother

\begin{document}

\date{}
\title{\Large\bf On Matrix Realizations of the Contact Superconformal Algebra $\hat{K}'(4)$ 
and the Exceptional $N = 6$ Superconformal Algebra}
\author{Elena Poletaeva\\[2mm]
School of Mathemtics, Institute for Advanced Study, 
Princeton and\\
 Department of Mathematics, University of Texas - Pan American (permanent address),\\
Email: elena@math.ias.edu and elenap@utpa.edu\\[2mm]
} \maketitle

{\footnotesize \noindent {\bf Abstract.} 
The superalgebra $\hat{K}'(4)$ and the exceptional $N = 6$ superconformal algebra have ``small''
irreducible representations in the superspaces $V^{\mu} = t^{\mu}\C[t, t^{-1}]\otimes\Lambda(N)$, where $N = 2$ and $3$, respectively.
For ${\mu \in \C\backslash \Z}$ they are associated to the embeddings of these superalgebras into
the Lie superalgebras of pseudodifferential symbols on the supercircle $S^{1|N}$.
In this work we describe  $\hat{K}'(4)$ and the exceptional $N = 6$ superconformal algebra
in terms of matrices over a Weyl algebra.
Correspondingly, we obtain realizations of their representations in $V^{\mu}$ for $\mu = 0$.
 } \vskip.2in

\section{Introduction}

This work  is a continuation of [20, 21].

Recall that {\it a superconformal algebra} is a simple complex Lie superalgebra spanned by the coefficients of a finite family of pairwise local fields 
$$a(z) = \sum_{n\in\Z}a_{(n)}z^{-n-1},$$
one of which is the Virasoro field $L(z)$ [3, 9--11]. 
Superconformal algebras play an important r\^{o}le in the string theory and conformal field theory.
They can also be described in terms of
derivations of the associative superalgebra 
$\C[t, t^{-1}]\otimes \Lambda(N)$, where $\Lambda (N)$ is the Grassmann algebra in $N$ variables.
The Lie superalgebra $K(N)$ of contact vector fields with Laurent polynomials as coefficients
is spanned by $2^N$ fields [3, 6, 7, 10].
It is also known as the $SO(N)$ superconformal algebra [1]. 
$K(N)$ is simple if $N \not= 4$, if $N = 4$, then the derived Lie superalgebra $K'(4)$ is simple.
The nontrivial central extensions of $K(1)$, $K(2)$, and $K'(4)$ are well-known:
they are isomorphic to the Neveu-Schwarz superalgebra, the ``$N = 2$'' superconformal algebra, and
the ``big $N = 4$'' superconformal algebra [1]. 
$K(6)$ contains the exceptional $N = 6$ superconformal algebra, also denoted by $CK_6$, 
as a subsuperalgebra. Note that $CK_6$ is  ``one half'' of  $K(6)$:
it is spanned by 32 fields [3, 4, 6, 12, 15, 22--24]. 

In [16, 17] Martinez and Zelmanov obtained $CK_6$ as a particular case of their construction of 
superalgebras $CK(R, d)$, where $R$ is an associative commutative
superalgebra with an even derivation $d$.

Our approach is based on the realization of $K(2N)$ in terms of pseudodifferential symbols on the circle extended by $N$ odd variables. It is well-known that a Lie algebra of contact vector fields can be realized
as a subalgebra of  Poisson algebra [2]. Analogously, $K(2N)$ can be embedded into 
the Poisson superalgebra $P(2N)$ of pseudodifferential symbols on the supercircle $S^{1|N}$ [20, 21].
There exists a family $P_h(2N)$ of Lie superalgebras of pseudodifferential symbols on $S^{1|N}$,
which contracts to $P(2N)$. There is no embedding of $K(2N)$ into $P_{h}(2N)$  if  $N\geq 3$.
It is remarkable that a nontrivial central extension $\hat{K}'(4)$ of $K'(4)$ and $CK_6$ can be embedded into $P_{h}(2N)$, where $N = 2$ and $3$, respectively [20, 21].

Associated to these embeddings, there are ``small'' irreducible representations of
$\hat{K}'(4)$ and $CK_6$ in the superspaces $V^{\mu} = t^{\mu}\C[t, t^{-1}]\otimes \Lambda(N)$, 
where $(\partial/{\partial t})^{-1}$ acts as an antiderivative.
This requires that $\mu\in\C\backslash\Z$.
Nevertheless,  the representations of $\hat{K}'(4)$ and $CK_6$ in $V^{\mu}$ can be defined if $\mu = 0$.
In this work we describe  these superalgebras in terms of matrices over  
the Weyl algebra $W = \sum_{i\geq 0}\A d^i$, where $\A = \C[t, t^{-1}]$ and $d = t\partial/{\partial t}$ 
(Theorems 1 and 2).
This gives realizations of the representations in $V^{\mu}$ for $\mu = 0$.

\section{Contact and Poisson superalgebras}

\noindent 
A {\it superconformal algebra} is a complex Lie superalgebra $\g$
such that

\noindent
(1) $\g$ is simple,

\noindent
(2) $\g$ contains the centerless Virasoro algebra $\hbox{der }\C [t, t^{-1}] = \oplus_{n\in \Z}\C L_n$
with the commutation relations
$$[L_n, L_m] = (m - n)L_{n+m} \eqno(1)$$
as a subalgebra,

\noindent
(3) $adL_0$ is diagonalizable with finite-dimensional eigenspaces:
$$\g = \oplus_i\g_i, \quad \g_i = \lbrace x\in\g \hbox{ }|\hbox{ } [L_0, x] = ix\rbrace, \eqno (2)$$
so that $\hbox{dim}\g_i<C$, where $C$ is a constant independent of $i$ [7].
\\[2mm]

\noindent
Let $\Lambda(2N)$ be the Grassmann algebra in $2N$ variables $\xi_1, \ldots, \xi_N$,
$\eta_1, \ldots, \eta_N$, and let 
$$\Lambda(1, 2N) = \C[t, t^{-1}]\otimes\Lambda(2N)$$
be the associative superalgebra with natural multiplication and with the following parity of generators:
$$p(t) = \bar{0},\hbox{ } p(\xi_i) = p(\eta_i) = \bar{1} \hbox{ for } i = 1, \ldots, N.$$
Let $W(2N)$ be the Lie superalgebra of all derivations of $\Lambda(1, 2N)$.
By definition,
$$K(2N) = \lbrace D\in W(2N) | D\Omega = f\Omega \hbox{ for some } f \in \Lambda(1, 2N)\rbrace$$
where $\Omega = \hbox{d}t + \sum_{i=1}^N(\xi_i\hbox{d}\eta_i + \eta_i\hbox{d}\xi_i)$
is a differential {\it contact} $1$-form [3--7, 10, 22--24].
There is a one-to-one correspondence between the differential operators
$D\in K(2N)$ and the functions $f\in \Lambda(1, 2N)$. Let 
$\partial_t, \partial_{\xi_i}$ and $\partial_{\eta_i}$ stand for $\partial/{\partial t}$,
$\partial/{\partial {\xi_i}}$ and $\partial/\partial {\eta_i}$, respectively.
The correspondence $f\leftrightarrow D_f$ is given by
$$D_f = \Delta (f)\partial_t + (\partial_tf)E - H_f,$$
where 
$$E = \sum_{i = 1}^N(\xi_i\partial_{\xi_i} + \eta_i\partial_{\eta_i}),\quad  \Delta = 2 - E,$$
$$H_f = (-1)^{p(f) + 1}\sum_{i = 1}^N(\partial_{\xi_i}f\partial_{\eta_i} + 
\partial_{\eta_i}f\partial_{\xi_i}).$$

\noindent The {\it Poisson algebra $P$ of pseudodifferential symbols on
the circle} is formed by the formal series 
$$A(t, \tau) = \sum_{-\infty}^na_i(t)\tau^i,$$
where $a_i(t)\in \C[t, t^{-1}]$, and the even variable $\tau$ corresponds to $\partial_t$.
The Poisson bracket is defined as follows:
$$\lbrace A(t, \tau), B(t, \tau)\rbrace = \partial_{\tau}A(t, \tau)\partial_tB(t, \tau) -
\partial_tA(t, \tau)\partial_{\tau}B(t, \tau).$$
An associative algebra
$P_h$, where $h\in (0, 1]$ is a deformation of $P$. The multiplication in $P_h$ is given as follows:
$$A(t, \tau)\circ_hB(t, \tau) = \sum_{n\geq 0 }{h^n\over {n!}}
\partial_{\tau}^nA(t, \tau)\partial_t^nB(t, \tau).$$
The Lie algebra structure on the vector space $P_h$ is given by 
$$[A, B]_h = A\circ_hB - B\circ_hA,$$
so that 
$$\hbox{lim}_{h\rightarrow 0}{1\over h}[A, B]_h = \lbrace A, B\rbrace, \eqno (3)$$
see [13, 14, 18, 19].
The {\it Poisson superalgebra of pseudodifferential symbols on $S^{1|N}$} is
$P(2N) = P\otimes \Lambda(2N)$. The Poisson bracket is defined as follows:
\begin{equation*}
\begin{aligned}
&\lbrace A, B\rbrace = \partial_{\tau}A\partial_tB - \partial_tA\partial_{\tau}B + \\
&(-1)^{p(A)+1}\sum_{i = 1}^N(\partial_{\xi_i}A\partial_{\eta_i}B + \partial_{\eta_i}A\partial_{\xi_i}B).
\end{aligned}
\end{equation*}
Let $\Lambda_h(2N)$ be an associative superalgebra with generators
$\xi_1, \ldots, \xi_N, \eta_1, \ldots, \eta_N$ and relations
$$\xi_i\xi_j = - \xi_j\xi_i,\quad \eta_i\eta_j = -\eta_j\eta_i, \quad \eta_i\xi_j = h\delta_{i, j} - \xi_j\eta_i.$$
Let $P_h(2N) = P_h\otimes\Lambda_h(2N)$ be a superalgebra with the product given by
$$(A_1\otimes X)(B_1\otimes Y) = (A_1\circ_hB_1)\otimes (XY),$$
 where
$A_1, B_1 \in P_h$ and $X, Y \in \Lambda_h(2N)$. 
The Lie bracket of $A = A_1\otimes X$ and $B = B_1\otimes Y$  is 
$$[A, B]_h = AB - (-1)^{p(A)p(B)}BA,$$
and (3) is satisfied.
$P_h(2N)$ is the {\it Lie superalgebra of pseudodifferential symbols on $S^{1|N}$}.
There exist embeddings of  $\hat{K}'(4)$ and $CK_6$
into $P_h(2N)$, where $N = 2$ and $N = 3$, respectively [20, 21].

\section{Case $\hat{K}'(4)$}
\noindent
The derived superalgebra 
$$K'(4) = [K(4), K(4)]$$
 is a simple ideal in $K(4)$
of codimension one, defined from the exact sequence 
$$0\rightarrow K'(4)\rightarrow K(4)\rightarrow \C D_{t^{-1}\xi_1\xi_2\eta_1\eta_2}\rightarrow 0.$$
The superalgebra $K'(4)$ is spanned inside $P(4)$ 
by the 12 fields:
\begin{equation*}
\begin{aligned}
&L_n = t^{n+1}\tau, \hbox{ } X_n^j = t^{n+1}\tau\xi_j,\hbox{ } Y_n = t^{n+1}\tau\xi_1\xi_2, 
\end{aligned}
\tag{4}
\end{equation*}
\begin{equation*}
\begin{aligned}
&L_n^i = t^{n}\eta_i,\hbox{ } X_n^{ji} = t^{n}\xi_j\eta_i, \hbox{ } Y_n^i = t^{n}\xi_1\xi_2\eta_i,\\
\end{aligned}
\tag{5}
\end{equation*}
where $i, j = 1, 2$,
and 4 fields
\begin{equation*}
\begin{aligned}
&F_n^0 = t^{n-1}\tau^{-1}\eta_1\eta_2,\\
&F_n^i = t^{n-1}\tau^{-1}\xi_i\eta_1\eta_2, \quad i = 1, 2,\\
&F_n^3 = t^{n-1}\tau^{-1}\xi_1\xi_2\eta_1\eta_2, \quad n\not= 0.
\end{aligned}
\end{equation*}
Note that $L_n$ is a Virasoro field [20, 21].
Let $\hat{K}'(4)$ be one of three independent central extensions of $K'(4)$, such that the
corresponding 2-cocycle  is 
\begin{equation*}
\begin{aligned}
&c(L_n, F_k^3) = \delta_{n+k,0},\quad n\not = 0,\\
&c(X_n^i, F_k^j) = (-1)^j\delta_{n+k,0}, \quad 1\leq i\not= j\leq 2,\\
&c(Y_n, F_k^0) = \delta_{n+k,0}.
\end{aligned}
\end{equation*}
The superalgebra $\hat{K}'(4)\subset P_h(4)$ is spanned by the 12 fields (4)--(5) and
4 fields:
$$F_{n, h}^0 = \tau^{-1}\circ_ht^{n-1}\eta_1\eta_2,\eqno (6)$$
$$F_{n, h}^i = \tau^{-1}\circ_ht^{n-1}\eta_1\eta_2\xi_i, \quad i = 1, 2, \eqno (7)$$
$$F_{n, h}^3 = \tau^{-1}\circ_ht^{n-1}\eta_1\eta_2\xi_1\xi_2 + {h\over n}t^n, \quad n\not= 0,\eqno (8)$$
and the central element $h\in P_h(4)$, so that
$$\hbox{lim}_{h\rightarrow 0}\hat{K}'(4) = K'(4)\subset P(4).$$ 

Let $V^{\mu} = t^{\mu}\C[t, t^{-1}]\otimes\Lambda(\xi_1, \xi_2)$, where $\mu\in\C\backslash\Z$.
We fix $h = 1$, and define
a representation of $\hat{K}'(4)$ in $V^{\mu}$ accordingly to the formulas (4)--(8). Namely,
$\xi_i$ is the operator of multiplication in $\Lambda(\xi_1, \xi_2)$,
$\eta_i$ is identified with $\partial_{\xi_i}$,  $\tau^{-1}$ is identified with an antiderivative,
and the central element $1\in P_{h = 1}(4)$ acts by the identity operator.
Consider the following basis in $V^{\mu}$:
\begin{equation*}
\begin{aligned}
&v_m^0(\mu) = {1\over{m+\mu}}t^{m+\mu},\quad v_m^1(\mu) = t^{m+\mu}\xi_1,\\
&v_m^2(\mu) = t^{m+\mu}\xi_2, \quad v_m^3(\mu) = t^{m+\mu}\xi_1\xi_2, \hbox{ }m\in\Z.
\end{aligned}
\end{equation*}
Explicitly, the action of $\hat{K}'(4)$ on $V^{\mu}$ is given as follows:
\begin{equation*}
\begin{aligned}
&L_n(v_m^0(\mu)) = (n + m + \mu) v_{m+n}^0(\mu),\\
&L_n(v_m^i(\mu)) = (m + \mu) v_{m+n}^i(\mu), \hbox{ } i = 1, 2, 3, \\
&X_n^i(v_m^0(\mu)) =  v_{m+n}^i(\mu),\hbox{ } i = 1, 2, \\
&X_n^1(v_m^2(\mu)) = (m+\mu)v_{m+n}^3(\mu), \\
&X_n^2(v_m^1(\mu)) = - (m+\mu)v_{m+n}^3(\mu), \\
&Y_n(v_m^0(\mu)) = v_{m+n}^3(\mu), \\
&L_n^i(v_m^i(\mu)) = (n + m + \mu) v_{m+n}^0(\mu),\hbox{ } i = 1, 2,\\
&L_n^1(v_m^3(\mu)) = v_{m+n}^2(\mu),\quad
L_n^2(v_m^3(\mu)) =- v_{m+n}^1(\mu),\\
&X_n^{ji}(v_m^i(\mu)) = v_{m+n}^j(\mu), \quad i, j = 1, 2,\\
&X_n^{ii}(v_m^3(\mu)) = v_{m+n}^3(\mu), \quad i, j = 1, 2,\\
&Y_n^i(v_m^i(\mu)) = v_{m+n}^3(\mu),\hbox{ }i = 1, 2,  \\
&F_{n, 1}^0(v_m^3(\mu)) = -v_{m+n}^0(\mu),\\
&F_{n, 1}^1(v_m^2(\mu)) = -v_{m+n}^0(\mu),\quad
F_{n, 1}^2(v_m^1(\mu)) = v_{m+n}^0(\mu),\\
&F_{n, 1}^3(v_m^i(\mu)) = {1\over n}v_{m+n}^i(\mu), \quad n\not= 0, 
\quad i = 0, 1, 2, 3.
\end{aligned}
\end{equation*}
Naturally,  $V^{\mu} = \oplus_m V^{\mu}_m$, where
$V_m^{\mu} = t^{m+\mu}\otimes \Lambda(\xi_1, \xi_2)$.
A $\Z$-grading on $\hat{K}'(4)$ is defined by the element $L_0 = t\tau$ of the Virasoro algebra
according to (2).
We have that
$$\g_i(V_m^{\mu}) \subset V_{m+i}^{\mu}, \eqno (9)$$
and $\g_0\cong \hat{\s\l}(2|2)$, where the central element is $L_0$.
Note that $\hat{\s\l}(2|2)$ has
the following one-parameter family $\hbox{spin}_{\lambda}$ of  
$(2|2)$-dimensional irreducible representations:
$$\hbox{spin}_{\lambda}:\left( \begin{array}{cc}
A & B \\
C & D 
\end{array}\right)\oplus \C L_0\rightarrow \left( \begin{array}{cc}
A & B + \lambda\tilde{C}\\
C & D
\end{array}\right)\oplus \C\lambda \cdot 1_{2|2},$$ 
where $A, B, C, D \in \g\l(2, \C)$, $trA = tr D$, $\lambda\in \C$.  
Let $E_{ij}$ be an elementary $2\times 2$-matrix. $\tilde{C}$ is determined by the following
conditions: 
\begin{equation*}
\begin{aligned}             
&\hbox{if } C = E_{ii}, \hbox{ then } \tilde{C} = E_{jj}, \hbox{ where } i\not= j,\\
&\hbox{if } C = E_{ij}, \hbox{ } i\not= j,
\hbox{ then } \tilde{C} = -E_{ij}.
\end{aligned}
\tag{10}
\end{equation*}
According to (9), there is a representation of  $\g_0$ in $V_m^{\mu}$ 
for each $m\in \Z$,
and   $V_m^{\mu}\cong \hbox{spin}_{\lambda = m+\mu}$ as $\g_0$-modules.

Note that if $\mu = 0$, we cannot formally define a representation of $\hat{K}'(4)$
in $V^{\mu}$. 
Nevertheless, all the formulas for the action of $\hat{K}'(4)$ on vectors $v_m^i(\mu)$, where $i = 0, 1, 2, 3$
and $m\in\Z$,
remain true to $\mu = 0$. Thus a representation of $\hat{K}'(4)$ in the superspace 
$$V = \hbox{Span}(v_m^i(0)\hbox{ } |\hbox{ } i = 0, 1, 2, 3 \hbox{ and } m \in \Z) \eqno (11)$$
 is well-defined.
To obtain a realization of this representation, at first
we will describe
$\hat{K}'(4)$ in terms of matrices over a Weyl algebra.
By definition, a Weyl algebra is
$$W = \sum_{i\geq 0}\A d^i, \eqno (12)$$
 where $\A$ is an associative commutative algebra and 
 
 \noindent
 $d:\A\rightarrow \A$ is a derivation of $\A$
 with the relations
$$da = d(a) + ad, \quad a\in\A.$$
Set 
$$\A = \C[t, t^{-1}], \hbox{ } d= L_0 = t\tau. \eqno (13)$$
Replacing $\lambda$ by $d$ in the formulas for $\hbox{spin}_{\lambda}$, we obtain the following theorem.

\noindent
{\bf Theorem 1.} Let $\hat{K}'(4) =  \oplus_{i}\g_i$, where the $\Z$-grading is given by 
$L_0 = t\tau$. Then

\noindent
$1)$ $\g_0\cong \hat{\s\l}(2|2)$ is realized as a Lie superalgebra of
$4\times 4$ matrices over $\C[d]$ of the type
$$\left( \begin{array}{cc}
A & B + d\tilde{C}\\
C & D
\end{array}\right)\oplus \C d\cdot 1_{2|2},$$ 
 where $A$, $B$, $C$, and $D$ are $2\times 2$ matrices over $\C$,
  $trA = trD$ and $\tilde{C}$ is determined by the conditions (10).
The central element in $\hat{\s\l}(2|2)$ is $L_0 = d\cdot1_{2|2}$, and the central element in 
$\hat{K}'(4)$ is $1_{2|2}$.

\noindent
$2)$ $\hat{K}'(4)$ is a subsuperalgebra of $4\times 4$ matrices over $W$ generated by
$\hat{\s\l}(2|2)$ and by all matrices 
$$\left( \begin{array}{cc}
E_{ij}(a) & 0\\
0 & 0
\end{array}\right), 
\left( \begin{array}{cc}
0 & 0\\
0 & E_{ij}(a)
\end{array}\right),$$ 
where $a\in \A$ and $1\leq i\not= j\leq 2$.

\noindent
$3)$ The standard representation of $\hat{K}'(4)$, realized as matrices over $W$, in $(2|2)$-dimensional vector superspace over $\A$
is isomorphic to the above-mentioned representation in the superspace $V$ 
in the case when $\mu = 0$, see (11).

\noindent
\section{Case $CK_6$}
\noindent
The exceptional superconformal algebra $CK_6$ is spanned by the following 
32 fields inside $K(6)\subset P(6)$:
\begin{equation*}
\begin{aligned}   
&L_n = t^{n+1}\tau, \quad G_n^i = t^{n+1}\tau\xi_i, \hbox{ } i = 1, 2, 3,\\
&\tilde{G}_n^i = t^n\eta_i - nt^{n-1}\tau^{-1}\xi_j\eta_i\eta_j, \quad i = 1, 2, 3,\\
&T_n^{ij} = t^n\xi_i\eta_j - nt^{n-1}\tau^{-1}\xi_k\xi_i\eta_k\eta_j, \quad i\not= j \not= k,\\
&T_n^i = -t^n(\xi_j\eta_j + \xi_k\eta_k) + nt^{n-1}\tau^{-1}\xi_j\xi_k\eta_j\eta_k, \quad i = 1, 2, 3,\\
&S_n^i = -t^n\xi_i(\xi_j\eta_j + \xi_k\eta_k) + nt^{n-1}\tau^{-1}\xi_i\xi_j\xi_k\eta_j\eta_k, 
\hbox{ } i = 1, 2, 3,\\
&\tilde{S}_n^i = t^{n-1}\tau^{-1}(\xi_j\eta_j - \xi_k\eta_k)\eta_i,\quad
 i = 1, 2, 3, 
\end{aligned}
\end{equation*}
\begin{equation*}
\begin{aligned}   
&I_n^i = t^{n-1}\tau^{-1}\xi_i\eta_j\eta_k, \quad i = 1, 2, 3, \quad
I_n = t^{n+1}\tau\xi_1\xi_2\xi_3,\\
&J^{ij}_n = t^{n+1}\tau\xi_i\xi_j, \quad \tilde{J}^{ij}_n = t^{n-1}\tau^{-1}\eta_i\eta_j, 
\quad i<j,
\end{aligned}
\end{equation*}
where $n\in\Z$, and $(i, j, k)$ is the cycle $(1, 2, 3)$ in the formulas for $\tilde G_n^i$, $T_n^i$,
$S_n^i$, $\tilde{S}_n^i$, and $I_n^i$, see [21].
Note that $L_n$ is a Virasoro field. 

$CK_6$ is spanned inside $P_h(6)$ by the 8 fields: $L_n$, $G_n^i$, $I_n$, and $J^{ij}_n$,
and the following 24 fields:
\begin{equation*}
\begin{aligned}             
&\tilde{G}_{n,h}^i = t^n\eta_i - n\tau^{-1}\circ_ht^{n-1}\eta_i\eta_j\xi_j, 
\hbox{ } i = 1, 2, 3,\\
&T_{n,h}^{ij} = t^n\xi_i\eta_j - n\tau^{-1}\circ_ht^{n-1}\eta_k\eta_j\xi_k\xi_i,  \hbox{ } i\not= j \not= k,\\
&T_{n,h}^i =  -t^n(\xi_j\eta_j + \xi_k\eta_k) + \\ &\qquad \quad n\tau^{-1}\circ_ht^{n-1}\eta_j\eta_k\xi_j\xi_k + ht^n,\\           
&S_{n,h}^i = -t^n\xi_i(\xi_j\eta_j + \xi_k\eta_k) + \\ &\qquad \quad n\tau^{-1}\circ_ht^{n-1}\eta_j\eta_k\xi_i\xi_j\xi_k + ht^n\xi_i,\\
&\tilde{S}_{n,h}^i = \tau^{-1}\circ_ht^{n-1}(\eta_j\eta_i\xi_j - \eta_k\eta_i\xi_k), 
\hbox{ }i = 1, 2, 3,\\
&I_{n,h}^i = \tau^{-1}\circ_ht^{n-1}\eta_j\eta_k\xi_i, \hbox{ } i = 1, 2, 3,\\
&\tilde{J}_{n,h}^{ij} = \tau^{-1}\circ_ht^{n-1}\eta_i\eta_j, 
\quad i<j, 
\end{aligned}
\tag{14}
\end{equation*}
where $n\in\Z$, and $(i, j, k)$ is the cycle $(1, 2, 3)$.
We have that $\hbox{lim}_{h\rightarrow 0}CK_6 = CK_6\subset P(6)$, see [21].

\noindent
Let $V^{\mu} = t^{\mu}\C[t, t^{-1}]\otimes\Pi(\Lambda(\xi_1, \xi_2, \xi_3))$, where $\mu\in\C\backslash\Z$,
and $\Pi$ denotes the change of parity.
We fix $h = 1$, and define
a representation of $CK_6$ in $V^{\mu}$ according to the formulas (14).
Consider the following basis in $V^{\mu}$:
\begin{equation*}
\begin{aligned}  
&v_m^i(\mu) = {t^{m+\mu}\over{m+\mu}}\Pi(\xi_i),\hbox{ }
\hat{v}_m^i(\mu) = t^{m+\mu}\Pi(\xi_j\xi_k), \hbox{ } 1\leq i\leq 3,\\
&v_m^4(\mu) = {t^{m+\mu}\over{m+\mu}}\Pi(1), \hbox{ } \hat{v}_m^4(\mu) = - t^{m+\mu}\Pi(\xi_1\xi_2\xi_3),
\end{aligned}
\end{equation*}
where $m \in \Z$ and $(i, j, k)$ is the cycle $(1, 2, 3)$ in the formulas for $\hat{v}^i_m(\mu)$.
Explicitly, the action of $CK_6$ on $V^{\mu}$ is given as follows:
\begin{equation*}
\begin{aligned}  
&L_{n}(v_m^i(\mu)) = (m + n + \mu)v_{m+n}^i(\mu),\\
&L_{n}(\hat{v}_m^i(\mu)) = (m + \mu)\hat{v}_{m+n}^i(\mu),\\
&G_{n}^i(v_m^4(\mu)) = (m + n + \mu)v_{m+n}^i(\mu),\\
&G_{n}^i(\hat{v}_m^i(\mu)) = -(m + \mu)\hat{v}_{m+n}^4(\mu),\\
&G_{n}^i(v_m^j(\mu)) = \hat{v}_{m+n}^k(\mu), \hbox{ } G_{n}^i(v_m^k(\mu)) = -\hat{v}_{m+n}^j(\mu),\\
&\tilde{G}_{n,1}^i(v_m^i(\mu)) = v_{m+n}^4(\mu), \hbox{ } \tilde{G}_{n,1}^i(\hat{v}_m^4(\mu)) = -\hat{v}_{m+n}^i(\mu),\\
&\tilde{G}_{n,1}^i(\hat{v}_m^j(\mu)) = -(m + \mu)v_{m+n}^k(\mu),\\
&\tilde{G}_{n,1}^i(\hat{v}_m^k(\mu)) = (m + n +\mu)v_{m+n}^j(\mu),\\
&T_{n,1}^{ij}(v_m^j(\mu)) = v_{m+n}^i(\mu), \hbox{ }
T_{n,1}^{ij}(\hat{v}_m^i(\mu)) = - \hat{v}_{m+n}^j(\mu),\\
&T_{n,1}^i(v_m^i(\mu)) = v_{m+n}^i(\mu), \hbox{ }
T_{n,1}^i(v_m^4(\mu)) = v_{m+n}^4(\mu),\\
&T_{n,1}^i(\hat{v}_m^i(\mu)) = -\hat{v}_{m+n}^i(\mu), \hbox{ }
T_{n,1}^i(\hat{v}_m^4(\mu)) = -\hat{v}_{m+n}^4(\mu),\\
&S_{n,1}^i(v_m^4(\mu)) = v_{m+n}^i(\mu),\hbox{ }
S_{n,1}^i(\hat{v}_m^i(\mu)) = \hat{v}_{m+n}^4(\mu),\\
&\tilde{S}_{n,1}^i(\hat{v}_m^j(\mu)) = v_{m+n}^k(\mu), \hbox{ }
\tilde{S}_{n,1}^i(\hat{v}_m^k(\mu)) = v_{m+n}^j(\mu),\\
&I_{n,1}^i(\hat{v}_m^i(\mu)) = - v_{m+n}^i(\mu), \hbox{ }
I_{n}(v_m^4(\mu)) = -\hat{v}_{m+n}^4(\mu),\\
&J_{n}^{ij}(v_m^4(\mu)) = \hat{v}_{m+n}^k(\mu), \hbox{ }
J_{n}^{ij}(v_m^k(\mu)) = - \hat{v}_{m+n}^4(\mu),\\
&\tilde{J}_{n,1}^{ij}(\hat{v}_m^k(\mu)) = -v_{m+n}^4(\mu), \hbox{ }
\tilde{J}_{n,1}^{ij}(\hat{v}_m^4(\mu)) = v_{m+n}^k(\mu),
\end{aligned}
\end{equation*}
where $(i, j, k)$ is the cycle $(1, 2, 3)$, see [21].

We have that  $V^{\mu} = \oplus_m V^{\mu}_m$, where
$V_m^{\mu} = t^{m+\mu}\otimes \Pi(\Lambda(\xi_1, \xi_2, \xi_3))$.
A $\Z$-grading in $CK_6$ is defined by the element $L_0 = t\tau$ of the Virasoro algebra
according to (2), so that (9) holds.
Note that $\g_0\cong \hat{\P}(4)$, where the central element is $L_0$ and
$\P (4)$  is a simple Lie superalgebra defined as follows.
Let $\tilde{\P}(4)$ be the Lie superalgebra, which preserves the odd nondegenerate supersymmetric
bilinear form $\hbox{antidiag} (1_4, 1_4)$ on the $(4|4)$-dimensional complex superspace.
Thus
$$\tilde{\P}(4) =\lbrace\left( \begin{array}{cc}
A & B \\
C & -A^t 
\end{array}\right) \hbox{ }|\hbox{ } A\in \g\l(4, \C), B^t = B, C^t = - C\rbrace.$$
$\P (4)$ is a subsuperalgebra of $\tilde{\P}(4)$ such that $A\in \s\l(4, \C)$, see [8].
$\hat{\P}(4)$ is a nontrivial central extension of $\P(4)$.
It is known that $\hat{\P}(4)$  has 
a family $\hbox{spin}_{\lambda}$ of $(4|4)$-dimensional irreducible representations:
$$\hbox{spin}_{\lambda}:\left( \begin{array}{cc}
A & B \\
C & -A^t 
\end{array}\right)\oplus \C L_0\rightarrow \left( \begin{array}{cc}
A & B - \lambda\tilde{C}\\
C & -A^t
\end{array}\right)\oplus \C\lambda \cdot 1_{4|4},$$ 
where $\lambda\in \C$, $1_{4|4}$ is the identity matrix, and $\tilde{C}$ is determined by the following
condition:
\begin{equation*}
\begin{aligned}  
&\hbox{if } C_{ij} = E_{ij} - E_{ji},
\hbox{ then } \tilde{C}_{ij} = C_{kl},
\hbox{ so that}\\
&\hbox{ the permutation } (1, 2, 3, 4) \mapsto (i, j, k, l) \hbox{ is even}, 
\end{aligned}
\tag{15}
\end{equation*} 
cf. [6] and [22--24].
According to (9), there is a representation of  $\g_0$ in $V_m^{\mu}$ for each $m\in \Z$,
and   $V_m^{\mu}\cong \hbox{spin}_{\lambda = m+\mu}$ as $\g_0$-modules.

Similarly to the case of $\hat{K}'(4)$,
all the formulas for the action of $CK_6$ on vectors $v_m^i(\mu), \hat{v}_m^i(\mu)$, 
where $1\leq i \leq 4$ and $m \in \Z$, remain true
to $\mu = 0$. Thus a representation of $CK_6$ in the superspace 
$$V = \hbox{Span}(v_m^i(0), \hat{v}_m^i(0)\hbox{ } |\hbox{ } 1\leq i \leq 4 \hbox{ and } m \in \Z)\eqno (16)$$
 is well-defined.
To obtain a realization of this representation,
we will use the Weyl algebra $W$ defined in (12) and (13).
Replacing $\lambda$ by $d$ in the formulas for $\hbox{spin}_{\lambda}$, we obtain the following theorem,
cf. [17]. 

\noindent
{\bf Theorem 2.}
 Let $CK_6 =  \oplus_{i}\g_i$, where the $\Z$-grading is given by 
$L_0 = t\tau$. Then

\noindent
 $1)$ $\g_0\cong \hat{\P}(4)$ is realized as a Lie superalgebra of
$8\times 8$ matrices over $\C[d]$ of the type
$$\left( \begin{array}{cc}
A & B - d\tilde{C}\\
C & -A^t
\end{array}\right)\oplus \C d\cdot 1_{4|4},$$
where $A$, $B$, and $C$ are $4\times 4$ matrices over $\C$, $tr A = 0$,
$B^t = B$, $C^t  = - C$, and $\tilde{C}$ is determined by the condition (15).
The central element in $\hat{\P}(4)$ is $L_0 = d\cdot1_{4|4}$.

\noindent
$2)$ $CK_6$ is a subsuperalgebra of $8\times 8$ matrices over $W$ generated by
$\hat{\P}(4)$ and by all matrices 
$$\left( \begin{array}{cc}
E_{ij}(a) & 0\\
0 & -E_{ji}(a)
\end{array}\right), \hbox{ where } a\in \A \hbox{ and } 1\leq i\not= j\leq 4.$$

\noindent
$3)$ The standard representation of $CK_6$, realized as matrices over $W$,
 in $(4|4)$-dimensional vector superspace over $\A$
is isomorphic to the above-mentioned representation in the superspace $V$
in the case when $\mu = 0$, see (16).

\section{Acknowledgements}

\noindent 
This material is based upon work supported by the National Science Foundation under agreement 
$No.\hbox{ } DMS-0111298$. Any opinions, findings and conclusions or recommendations expressed in this material are those of the author and do not necessarily reflect the views of the National Science Foundation.

The author is grateful to the 
Institute for Advanced Study for the hospitality and support during term II of the academic year 2006--2007.
She wishes to thank the organizers of the 5th International Conference on Differential Equations and Dynamical Systems. She is also grateful to 
V. Serganova for very useful discussions.

\footnotesize
\section{References}

\begin{itemize}
\item[{[1]}]
M. Ademollo, L. Brink, A. D'Adda et al., Dual strings with $U(1)$ colour symmetry, {\em Nucl. Phys.},
B {\bf 111}, (1976) 77--110.

\item[{[2]}]
V.I. Arnold, 
Mathematical Methods of Classical Mechanics,
Springer, New York, 1989.

\item[{[3]}]
S.-J. Cheng and V. G. Kac, A new $N = 6$ superconformal algebra,
{\em Commun. Math. Phys.}, {\bf 186}, (1997) 219--231.

\item[{[4]}]
S.-J. Cheng and V. G. Kac, Structure of some $\Z$-graded  Lie superalgebras of vector fields,
{\em Transform. Groups}, {\bf 4}, (1999) 219--272.

\item[{[5]}]
B. Feigin and D. Leites, New Lie superalgebras of string theories, in
{\em Group-Theoretical Methods in Physics}, edited by M. Markov et al.,
Nauka, Moscow {\bf 1}, (1983) 269--273.
(English translation Gordon and Breach, New York, 1984).

\item[{[6]}]
P. Grozman, D. Leites, and I. Shchepochkina,
Lie superalgebras of string theories,
{\em Acta Math. Vietnam.}, {\bf 26}, (2001) 27--63; hep-th/9702120.

\item[{[7]}]
V. G. Kac and J. W. van de Leur, On classification of superconformal algebras, in 
{\em Strings-88}, edited by S. J. Gates et al.,
world Scientific, Singapore, (1989) 77--106.

\item[{[8]}]
V. G. Kac, Lie superalgebras,
{\em Adv. Math.}, {\bf 26}, (1977) 8--96.

\item[{[9]}]
V. G. Kac, Classification of supersymmetries,
{\em Proceedings of the International Congress of Mathematicians}, {\bf 1}, Beijing (2002) 319--344,
Higher Ed. Press, Beijing, 2002.

\item[{[10]}]
V. G. Kac, Superconformal algebras and transitive group actions on quadrics,
{\em Commun. Math. Phys.}, {\bf 186}, (1997) 233--252. Erratum: {\bf 217}, (2001) 697--698.

\item[{[11]}]
V. G. Kac, 
Vertex Algebras for Beginners,
University Lecture Series, Vol. 10, AMS, Providence, RI, 1996.
Second edition, 1998.

\item[{[12]}]
V. G. Kac, Classification of infinite-dimensional simple linearly compact Lie superalgebras,
{\em Adv. Math.}, {\bf 139}, (1998) 1--55.

\item[{[13]}]
B. A. Khesin, V. Lyubashenko, and C. Roger,
Extensions and contractions of the Lie algebra of
$q$-pseudodifferential symbols on the circle,
{\em J. Funct. Anal.}, {\bf 143}, (1997) 55--97.

\item[{[14]}]
O. S. Kravchenko and B. A. Khesin,
Central extension of the algebra of pseudodifferential symbols,
{\em Funct. Anal. Appl.}, {\bf 25}, (1991) 83--85.

\item[{[15]}]
D. Leites and I. Shchepochkina,
preprint MPIM2003--28.

\item[{[16]}]
C. Martinez and E. I. Zelmanov,
Simple finite-dimensional Jordan superalgebras of prime characteristic,
{\em J. Algebra}, {\bf 236}, (2001) 575--629.

\item[{[17]}]
C. Martinez and E. I. Zelmanov,
Lie superalgebras graded by $P(n)$ and $Q(n)$,
{\em Proc. Natl. Acad. Sci. USA}, {\bf 100}, (2003) 8130--8137.

\item[{[18]}]
V. Ovsienko and C. Roger, 
Deforming the Lie algebra of vector fields on $S^1$ inside the Poisson algebra on $\dot{T}^*S^1$.
{\em Commun. Math. Phys.}, {\bf 198}, (1998) 97--110.

\item[{[19]}]
V. Ovsienko and C. Roger,
Deforming the Lie algebra of vector fields on $S^1$ inside the Lie algebra
 of pseudodifferential symbols on $S^1$,
{\em Amer. Math. Soc. Transl.}, {\bf 194}, (1999) 211--226.

\item[{[20]}]
E. Poletaeva,
A spinor-like representation of the contact superconformal algebra $K'(4)$,
{\em J. Math. Phys.}, {\bf 42}, (2001) 526--540.

\item[{[21]}]
E. Poletaeva,
On the exceptional $N = 6$ superconformal algebra,
{\em J. Math. Phys.}, {\bf 46}, (2005) 103504, 13 pp.
Publisher's note, {\em J. Math. Phys.}, {\bf 47}, (2006) 019901, 1 p.

\item[{[22]}]
I.  Shchepochkina, hep-th/9702121.

\item[{[23]}]
I. Shchepochkina,
Five simple exceptional Lie superalgebras of vector fields,
{\em Funktsional. Anal. i Prilozhen.}, {\bf 33}, (1999) 59--72.
Translation in {\em Funct. Anal. Appl.}, {\bf 33}, (1999) 208--219.

\item[{[24]}]
I. Shchepochkina,
The five  exceptional simple Lie superalgebras of vector fields and their fourteen regradings,
{\em Represent. Theory}, {\bf 3}, (1999) 373--415.

\end{itemize}
\end{document}